# Immersive VR as a Tool to Enhance Relaxation for Undergraduate Students with the Aim of Reducing Anxiety – A Pilot Study

**J Lewis, B Rorstad**

Department of Computing, Nottingham Trent University

Corresponding author: J. Lewis  (e-mail: james.lewis@ntu.ac.uk).

**ABSTRACT** Despite extensive use in related domains, Virtual Reality (VR) for generalised anxiety disorder (GAD) has received little previous attention.  We report upon a VR environment created for the Oculus Rift and Unreal Engine 4 (UE4) to investigate the potential of a VR simulation to be used as an anxiety management tool. We introduce the broad topic of GAD and related publications on the application of VR to this, and similar, mental health conditions. We then describe the development of a real time simulation tool, based upon the passive VR experience of a tranquil, rural alpine scene experienced from a seated position with head tracking.  Evaluation focused upon qualitative feedback on the application. Testing was carried out over the period of two weeks on a sample group of eleven students studying at Nottingham Trent University. All participants were asked to complete the Depression, Anxiety and Stress Scale – 21 Items (DASS21) at the beginning and at the end of the study order to assess their profile, and hence suitability to comment upon the software. Qualitative feedback was very encouraging, with all participants reporting that they believed the experience helped and that they would consider utilising it if it was available.  Additionally, a psychologist was asked to test the application to provide a specialist opinion on whether it would be appropriate for use as an anxiety management tool. The results highlight several areas for improvement but are positive overall in terms of its potential as a therapeutic tool.

## I. INTRODUCTION

Generalised Anxiety Disorder (GAD) is a disorder of chronic uncontrollable worry, with a lifetime population prevalence of around 5% [1]. It is often considered that GAD is not as serious as other anxieties, such as social anxiety or specific phobias and research shows that it is not uncommon for patients to wait for a considerable number of years before seeking treatment [2]. However, GAD can have a significant impact upon a person's life due to its long-term and chronic nature. Anxiety does not simply impact upon the psychosocial conduct of the affected individual, there are physiological symptoms such as difficulty concentrating, muscle tension and disturbed sleep. It has also been shown to have in impact upon mortality with increased risk of suicide [3] and death from cancer[4]. The development of effective treatment regimens for generalised anxiety is therefore a matter of substantial interest and importance to public health and wellbeing.

People usually develop GAD from the late teens and early twenties onwards and it typically lasts for around 20 years [1]. It is also not uncommon for people diagnosed with GAD to experience other types of anxiety or mood disorders. These include social anxiety, depression or alcohol misuse and these are often experienced at the same time as the GAD [1].People tend to seek treatment for these problems individually, rather than for their tendency to worry excessively [3].

VR is a powerful technology which has proven itself to be a potentially effective way to enhance psychotherapy [5]. The aim of this project was therefore to investigate whether healthcare for young adults with GAD can be enhanced by using VR.

## II. BACKGROUND AND EXISTING TREATMENT REGIMES

Multiple studies have been conducted regarding the use of VR to treat anxiety disorders through forms of VR exposure therapy (VRET). These include flying [6]-[8], spiders [9], [10], heights [11], public speaking [12], agoraphobia [13], [14] ,and post-traumatic stress disorder [15]. In general, these interventions have been found to be effective and acceptable to users. Negative side effects are not generally serious and in most cases factors such as disorientation and motion sickness can be associated with shortcomings in the VR technology available rather than the method per se.

Despite there being a large body of evidence supporting the effectiveness of VR in the treatment of specific anxiety disorders, research on the use of VR applications in the treatment of GAD is still at a very early stage. GAD has been well documented as one of the most difficult anxiety disorders to treat, yielding lower treatment response relative to other anxiety disorders [16]. It would therefore be of substantial interest if it can be shown that VR can play an effective role in this context.

It has been shown that relaxation is an effective approach in the treatment of GAD [17]. One of the limitations however, is that, for somebody who is anxious, depressed or stressed, a state of relaxation can be difficult to achieve. To overcome this problem, VR has been used as a means of facilitating the relaxation process by visually presenting key relaxing images to the patients. Gorini and Riva [18] suggested that VR imagery makes the experience more "real" than most people can create using their own imagination and that VR therefore induces a sense of presence which can enhance the relaxation experience. The randomised controlled trial of Gorini et al [19] combined a VR headset with a mobile phone, facilitating the relaxation therapy in an outpatient setting, albeit not in an immersive fashion. It was shown that this use of VR for relaxation represents a promising approach when it comes to treating individuals diagnosed with GAD. The perceived sense of presence was cited as a factor; however, the results also suggest that a mobile phone lacking in immersive content can still be useful to help individuals relax in an outpatient setting, with 91% of the participants reporting satisfaction with using a mobile phone to practice relaxation exercises at home.

Baños et al. [20], investigated stress-related disorders. Examining the efficacy of using a VR application for treating stress-related disorders, including post-traumatic stress disorder (PTSD), pathological grief (PG) and adjustment disorders (AD). The study indicated that cognitive behavioural therapy (CBT) with the addition of the virtual environments was at least as effective as the traditional CBT therapy. Furthermore, where significant differences existed these were in favour of the VR therapy. The data obtained in this study showed that VR can be used to improve a patient's relaxation intensity, thereby justifying continued investigation.

## III. AIMS

The aim of the project is to develop a VR application which could be used as a tool to reduce stress and anxiety amongst young adults with GAD. The application aims to facilitate relaxation by placing the user in a naturalistic VR environment which can simply be experienced. The user has no tasks to perform or demands made upon them. The intentions at this stage were to make a quick assessment about the usability of the technology utilised, the visual fidelity possible and the acceptability amongst the user group of this kind of intervention.

Previous attempts at using VR to treat GAD included some kind of interactivity allowing the user to move around in a three dimensional scene using a head-mounted display (HMD) and a keyboard [18], [19], [21]. Whilst the relationship between interactivity and presence is well established [22] it is unclear whether deep presence is a necessary component in VR relaxation, especially when considering Gorini's [19] utilisation of imagery on a mobile phone. Whilst a room scale six degree of freedom experience is well within the capabilities of the existing desktop systems, these have substantial limitations in terms of convenience and widespread access. Standalone mobile solutions, such as Gear VR or Oculus GO offer great potential in terms of low cost, ease of use and 'low friction' in terms of adoption, but to do so they sacrifice the translational degrees of freedom, making exploration of an environment more difficult to facilitate in a naturalistic way.

## IV. METHODS

*A. DESIGN*

The starting point of the research was the idea that naturalistic environments were supportive of relaxation and contemplation. There remained a specific question over exactly what kind of environment should be developed. To help inform this stage of the design a cohort of students in the Computing Department of Nottingham Trent University were polled with the question;

*"If you could imagine yourself in a peaceful and safe environment – a place that makes you relaxed and happy, where would it be?"*

| Answer Choices | Responses | |
|---|---|---|
| A beach | 35.48% | 33 |
| A forest | 10.75% | 10 |
| A winter environment | 8.60% | 8 |
| By a lake | 45.16% | 42 |
| Total | | 93 |

As the lake had the greatest support this was the starting point for the design. It was recognised that the context of an alpine lake would facilitate the incorporation of the 'Happy Place' proposed triggers of snow and trees in an attempt to maximise the reach of a single simulated environment.

The environment was developed using the Unreal Engine 4 game engine by Epic Games, with assets including trees, plants and flowers modelled in 3Ds Max and supplemented with assets already available from Epic Games in the marketplace. Custom textures for the landscape and water shaders were developed using Adobe Photoshop. Spatially referenced sound loops of running water, wind in the trees, and birdsong were also incorporated. Figures 1 and 2 illustrate the graphical fidelity achieved and show the bench which was the locus of the user's experience.

The low-cost Oculus based Samsung GearVR was initially selected as the development platform. The device costs around £50 and is therefore a relatively inexpensive platform, a key factor when considering the potential for widespread adoption. It is also the bestselling advanced consumer VR device of all time with over 5 million units in circulation [23]. This platform was therefore a mature and viable test bed for a potential rollout of the system. As the prototype evolved it became clear that the rendering requirements for the detailed naturalistic landscape planned was exceeding the capabilities of the Samsung smartphone which was powering the system. Rather than simplify the environmental features a decision was made to utilise the Oculus Rift instead. It is also worth noting that the development methodology employed, in particular the utilisation of the Unreal 4 Game engine, means that the application developed is somewhat platform agnostic. Whilst it was necessary to utilise a desktop PC at this time, for this phase of the work, it was clear that faster phones & improved mobile VR systems such as the HTC Vive Focus or Oculus Quest would soon become viable platforms for rich and detailed VR experiences. Redeploying the software prototype from a desktop implementation to mobile VR would be a relatively trivial step as the prototype remained within the expected operating parameters of newer mobile VR devices.

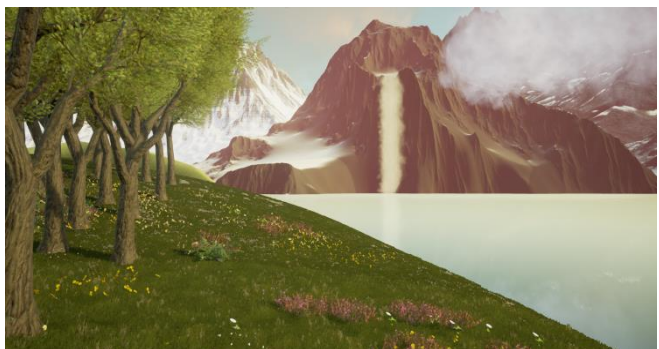

**Figure 1 - The simulated alpine scene**

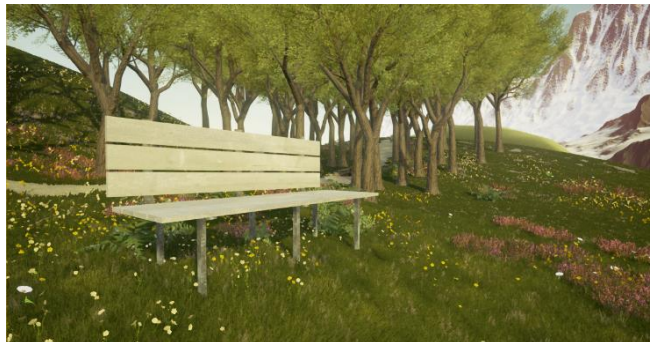

**Figure 2 - The Bench**

*B. Testing Tools & Methods*

The study procedures were conducted in accordance with the Declaration of Helsinki and received ethical approval from the Research Ethics Committee of the authors' university. All participants provided written informed consent prior to their participation in this study.

Participants were a mixed sex group of volunteer undergraduate students aged 20-22. To get an understanding of each participant's level of anxiety, all participants were asked to complete the Depression, Anxiety and Stress Scale – 21 items (DASS21) both at the start of the user testing and two weeks later, at the end of the user testing. The DASS21 uses a 4-point scale which measures the frequency or severity of the participants' experiences over the last week. The scale is a great tool for separating out the analysis across depression, anxiety and stress [24]. Our aim in applying this scale was to provide a glimpse at the mental health profile of our undergraduate participants. Participants volunteered for the trial and there was no attempt to profile or manage recruitment towards a representative sample. The study was advertised as a trial of a virtual reality experience and was not presented to potential volunteers as a mental health intervention at the outset. The sample is therefore likely to be biased towards individuals with an interest in technology, but not necessarily those seeking help with their mental health.

All participants were given a short explanation of how the equipment works, with an introduction stating that they experience an alpine scene showing a calm lake surrounded by mountains together with sounds such as birds tweeting and running water. Participants were instructed to look around to observe the surroundings whilst remaining seated on a bench and to attempt to simply relax. The participant was then seated on an office chair, assisted in donning the Oculus Rift headset and positioning the headphones. Testing was based on two 15-minute sessions. The sessions were separated by a duration of two weeks due to participant availability and to facilitate a period of reflection.

## V. RESULTS AND EVALUATION

### A. Quantitative Results

The Quantitative results obtained from the DASS21 questionnaire are presented first to illustrate the profile of the participants. Changes in the scale are discussed where noted, however the authors make no claim of significance associated with these, they are merely discussed for completeness and as preliminary indicators only. In interpreting the DASS21 the conventional cut-off values for severity thresholds indicated by Lovibond & Lovibond have been utilised [25]

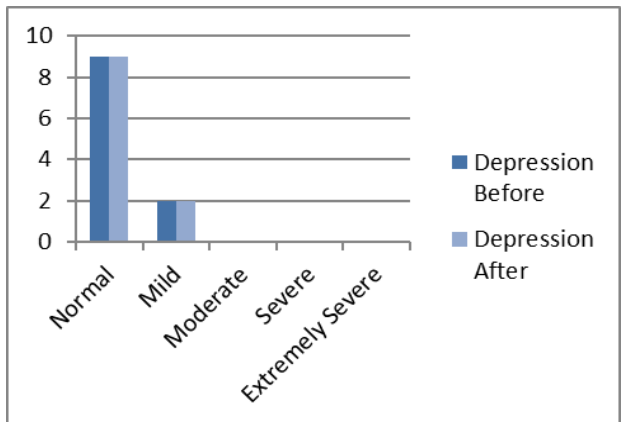

Figure 3 - Level of depression before and after testing

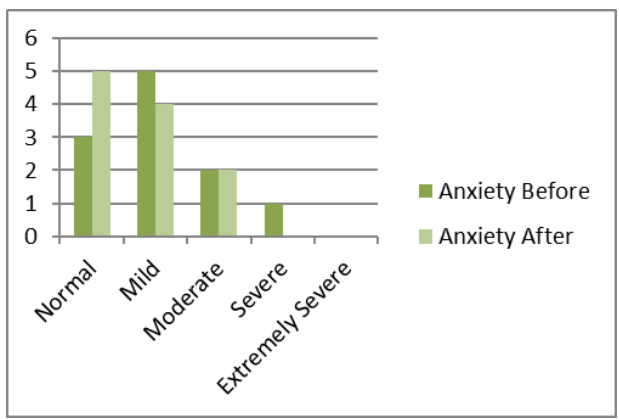

Figure 4 - Level of anxiety before and after testing

The key observations of relevance to this study are the results relating to stress and anxiety. It is interesting to note that only three participants indicated normal levels of stress and only three indicated normal levels of anxiety. All other indications are of elevated levels of stress and anxiety. The authors present this to indicate that the test group are suitably representative candidates to comment upon the application. It is also interesting to note the extent of stress and anxiety amongst undergraduate students, something which many academic staff will have anecdotally observed but which is laid bare by these figures

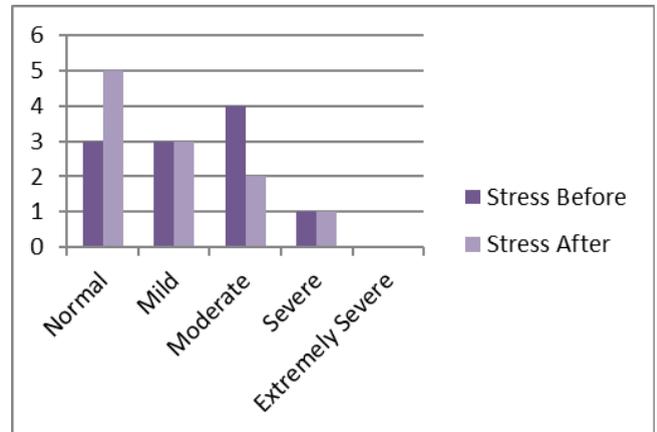

Figure 5 - level of stress before and after testing

The authors also suggest these results indicate that the application appears to cause no harm, and that there are tentative indications that it might help reduce levels of stress and anxiety. More robust methods, including a 'waiting list' control condition and a larger sample will be employed in future testing to see if any statistical significance underlies these observations.

### B. Qualitative Results

Turning to the qualitative results, which were the key purpose of this phase of the research, we asked three questions. These were designed to test assumptions made at the outset, and reported in the literature, as well as indicate the desirability of this type of intervention. A five point Likert scale was utilised and the questionnaire presented immediately after the second experience of the VR environment.

*I felt like the application helped me relax better than if I were to use my own imagination.*

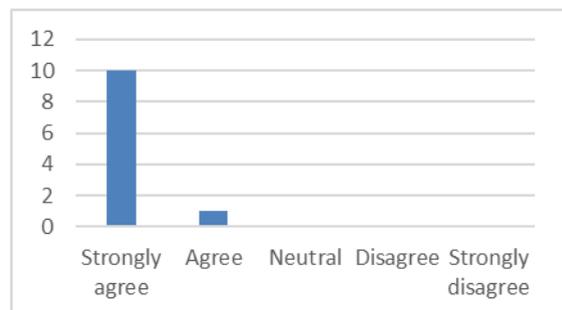

Figure 6 - Relaxation effectiveness

This question was designed to elicit an individual perspective on how the participant *felt* with regard to the utility of the application. Figure 6 shows that all participants believe the simulation helps them to relax more effectively than they could through simply applying their own imagination. The authors take this belief as an indication that participants are likely to have faith in the application, and are therefore likely to remain engaged for the duration of a longitudinal study. Ten out of eleven participants strongly agreed with the statement, this was a stronger reaction than the authors anticipated and indicates that for these users at least, they believe the application to be an effective tool to enhance relaxation.

It is important to note that participants were not asked to attempt to utilise their imagination as a relaxation mechanism as a formal part of the study. This question therefore does not objectively tell us whether or not the application is more effective than imagination alone could be. It is unlikely that the participants will have have been trained to utilise their imagination effectively in this context, and it is very possible that with such guidance this result could look very different.

*If given the opportunity, I would consider using this application as a tool to help me relax when feeling stressed or anxious.*

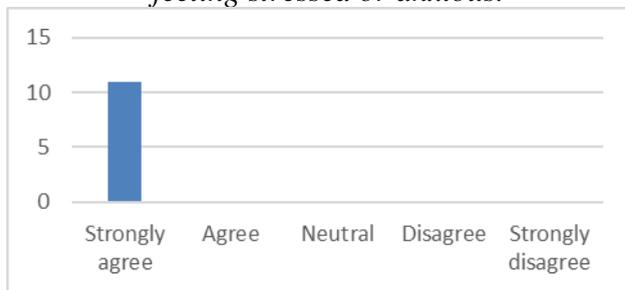

Figure 7 - User acceptance of the tool

As can be seen from figure 7, every single participant reported that they would consider using the application as a tool for enhancing relaxation when feeling stressed or anxious. This suggests that if it could be made easily available, in a "low friction" manner, the application could be attractive to a large number of people. The authors did not anticipate such a strongly positive response to this question but note that there is a difference between "considering the use of" something and stating that you "would actually use" something. It would be useful to investigate what, if any, factors convert somebody considering use of the VR simulation into somebody who does or doesn't actually use the tool. This will form the qualitative focus of future studies.

*I felt a strong sense of presence in the environment.*

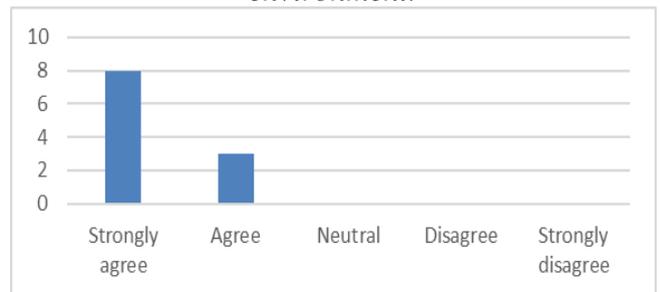

Figure 8 - Sense of presence

The results shown in figure 8 show that all the participants of the study felt a sense of presence in the virtual environment. This was in line with the anticipated result. The restriction of the participant to a seated position may have led to a reduced sense of presence, versus experiencing at room scale. However, the authors present this as indicative that 3 degree of freedom mobile VR solutions are likely of offer a degree of presence sufficient for effective VR relaxation therapy.

In order to identify any potential improvements to the application, participants were encouraged to give feedback on the system throughout the course of the user testing.

*"It felt really weird when I looked down and couldn't see my body. I don't know if would help, but I think it would be better if you could see a body instead of being invisible."*

Several participants commented on this theme, highlighting that the absence of an avatar may lower the sense of presence within the environment. An appropriate mechanism for achieving this for a diverse group of users, thereby enabling the participant to identify as the avatar requires significant further research.

*"I am currently on the waiting list to receive CBT treatment as I have just been diagnosed with generalised anxiety disorder… I've been told that the waiting time is approximately eight weeks… I feel like this app could really help me relax more whilst waiting for treatment... Although it wouldn't necessarily make me feel that different, I'm sure it would make more of a difference than not doing anything at all."*

This chance anecdote from within a small sample is perhaps indicative of the prevalence and individual experience of young people with GED. The authors also noted that in this case the participant indicates that they would use it, even though they did not anticipate it making much difference. Such an observation perhaps reveals a kind of desperation for anything to fill the treatment void in which people find themselves. Even where the perceived chance of success seems slim. It also cautions against making the leap from something people say they like, and report helps them relax, to making the assumption that it therefore might provide meaningful therapeutic benefit.

## C. EXPERT REVIEW
An interview with a professional psychologist was conducted in order to assess the potential use of the application. The psychologist was given a short explanation of how the system worked before testing it.

The psychologist identified that the system only consisted of one environment.

"*The environment looks great, but you have to remember that people are different and should therefore be provided with different options as to what they find relaxing.*"

This highlights that the application could benefit from having more than one environment. Although results indicate that the lake environment seems to be effective in releasing stress and anxiety, the application could have a bigger impact if more environments were added for the user to choose from.

The virtual environment was described as looking very realistic. However, one suggestion was made that could potentially increase the sense of presence within the environment.

"*The scene looks very realistic and I had a strong sense of presence. I don't know if it's possible, but it would be good to be able to walk around the environment as well as this could potentially improve the sense of presence.*"

Although this highlights an area for improvement, movement around the environment was not encouraged at this stage due the potential risk of trips or falls and due to a desire to mirror the experience possible through widely accessible mobile VR.

"*The birds tweeting and the sound of running water make the environment seem more realistic, but I feel like the application would be more useful for relaxation if a voice over was added in order to guide the patient in what to do.*"

This highlights that although there is a strong sense of presence, the patients might find the application more useful if guided through the relaxation experience.
After assessing some areas of the system, the psychologist was asked two fixed questions in order to assess the usefulness of the application.

***Do you think GAD patients could potentially benefit from this application?***

"*Yes, absolutely. Although it would be better to have more options to choose from in terms of environments, I think the app shows great potential and would be of great help especially if used together with therapy.*"

**If given the opportunity, would you consider using this application on some of your patients?**

"*I don't know much about VR to be honest, so I would have to look into it in a bit more detail first, but yes, I would definitely consider it as this seems like a great way of enhancing the relaxation experience.*"

It is encouraging to note that there is an openness and enthusiasm for VR from practitioners, even where their prior experience has not exposed them to state of the art practice in the use of VR for psychotherapies. This openness implies that barriers to adoption are not actually particularly significant, and that widespread adoption could quickly follow proof of clinical effectiveness.

## VI. Conclusions
Generalised anxiety disorder has been shown to be a widespread and pernicious disease which substantially impacts upon the lives of people. The high prevalence of the disorder, and the absence of effective and accessible treatment regimens makes the investigation of novel approaches to managing the condition one of significant interest.

The Unreal 4 computer game engine, together with 3D design tools can be utilised to create a synthetic environment with a high degree of realism. When combined with an Oculus Rift head mounted display the fidelity of the experience was sufficient to evoke a sense of presence in all users. The reported level of presence is high despite the interactive parameters of the simulation being restricted to those matching the technical capabilities of mobile VR solutions.

All participants reported that the simulation was helpful in facilitating relaxation, and every participant who experienced the simulation reported that they would consider using the tool to help them relax it if were made available for them to use. It was reviewed by a domain specialist who made several suggestions for future improvements, but who was overall supportive of the potential offered by the application.

This study provided encouraging evidence supporting the implementation of a larger scale trial. It is currently planned that this trial will offer a selection of naturalistic environments to better reflect the preferences of individual participants. It will also be delivered via a mobile, wireless headset, facilitating a lower friction route to wider adoption. Evaluation will be focused upon increasing the confidence measures of the preliminary quantitative results reported here through a larger scale randomised trial incorporating a waiting list control condition. It will also incorporate the use of EEG methods as an objective comparison to the self-reported measures of relaxation effectiveness.